\documentclass{epl}

\usepackage{bbm}
\usepackage{epsfig}
\usepackage{graphicx}
\usepackage[centertags]{amsmath}
\usepackage{amssymb}
\usepackage{amsfonts}
\usepackage{psfrag}

\title{Goldstone fluctuations in the amorphous solid state}

\author{
Swagatam Mukhopadhyay\inst{1}
\and 
Paul M.~Goldbart\inst{1} 
\and
Annette Zippelius\inst{2}}
\shortauthor{S. Mukhopadhyay \etal}

\institute{
\inst{1}Department of Physics, 
University of Illinois at Urbana-Champaign,\\ 
1110 West Green Street, Urbana, Illinois 61801-3080, U.S.A.\\
\inst{2}Institut f\"ur Theoretische Physik,
  Georg-August-Universit\"at G\"ottingen,\\ 
  37073 G\"ottingen, Germany}

 \pacs{61.43.-j}{Disordered solids}
 \pacs{82.70.Gg}{Gels and sols}

\def\wh{\widehat}
\def\hrs{{\rm HRS}}
\def\lpc{\ell_{\rm p}}
\def\fim{{\cal D}}
\def\rpf{{\cal Z}}
\def\ers{{\epsilon}}
\def\stiff{{\rho}}
\def\dbar{{\mathchar'26\mkern-12mud}}   
\def\newleta{{l_{<}}}

\begin{document}

\maketitle

\begin{abstract}
Goldstone modes in the amorphous solid state, resulting from the 
spontaneous breaking of translational symmetry due to random 
localisation of particles, are discussed.  
Starting from a microscopic model with 
quenched disorder, the broken symmetry is identified to be that of 
relative translations of the replicas.  Goldstone excitations, 
corresponding to pure shear deformations, are constructed from long 
wavelength distortions of the order parameter.  The elastic free 
energy is computed, and it is shown that Goldstone fluctuations 
destroy localisation in two spatial dimensions, yielding a 
two-dimensional amorphous solid state characterised by power-law 
correlations.
\end{abstract}

The {\it elastic\/} properties of gels and vulcanised matter have
received considerable attention in the literature. Phenomenological
arguments\cite{Treloar} as well as microscopic
models\cite{Castillo+Goldbart1998-2000} 
have been used to compute, {\it e.g.\/}, 
the shear modulus.  The {\it structural\/} properties of gels have been
discussed more recently\cite{GoldbartAdvPhy1996}, building on the work
of Edwards and collaborators \cite{Deam+Edwards1975}. It has been shown
that in the gel phase a finite fraction of particles is localised
around random positions with finite thermal excursions.  What has not
been analysed so far, and what constitutes the main focus of this
Letter, is the connection between the spontaneously broken
translational symmetry due to particle localisation and shear
deformations, which are identified with Goldstone excitations as a
consequence of the spontaneous breaking of translational symmetry.

We first discuss symmetry-related equilibrium states in the general
context of amorphous solids. We construct low-energy long-wavelength
Goldstone fluctuations, which correspond to almost-uniform translations
of the localised particles. Particular care is taken of the residual
symmetry of the amorphous solid state, which remains statistically
homogeneous. We then apply these general ideas to a particular type of 
amorphous solid---gels and vulcanised matter---for which a replica
field theory is available\cite{GoldbartAdvPhy1996}. 
This allows us to identify the lost symmetry 
to be that of relative translations of the replicas, whilst common
translations of all replicas remain as intact symmetries in the
amorphous solid state.  Goldstone fluctuations are described as
long-wavelength distortions of the order parameter, which are
reminiscent of the ripples of an interface between coexisting 
liquid and gas phases.  The free energy of these fluctuations is 
computed within a Gaussian approximation, and identified with the 
elastic free energy of 
a macroscopically homogeneous isotropic amorphous solid.

\section{Symmetry breaking and Goldstone modes}
We consider a system of $N$ distinguishable particles in a volume $V$
interacting {\it via\/} potentials that are 
translationally invariant, {\it e.g.\/}, 
additive pair potentials.  In the
amorphous solid phase a finite fraction of the particles are localised
at random positions with thermal fluctuations of finite range. The
system will be found in one of many possible equilibrium states. One
such state is singled out arbitrarily, denoted by $\nu$ and specified
by nonzero local static density fluctuations 
$\langle{\rm e}^{i{\bf k}\cdot{\bf R}_{j}}\rangle_{\nu}$. 
Here, ${\bf R}_{j}$ is the position of particle $j$ and $\langle\cdots\rangle_{\nu}$ denotes a thermal average
restricted to equilibrium state $\nu$.  The amorphous state has
macroscopic translational invariance, so that
\begin{equation}
\label{density}
\lim_{N\to \infty}
        \left[
\frac{1}{N}\sum\nolimits_{j=1}^{N}
\langle{\rm e}^{i{\bf k}\cdot{\bf R}_{j}}\rangle_{\nu}
        \right]
=0, 
\end{equation}
where $[\cdots]$ denotes a disorder average. 
Localisation in the equilibrium state $\nu$ is detected by 
higher-order moments, the simplest one being the second:
\begin{equation}
\label{naive_orderpar}
\Omega_2({\bf k}_{1},{\bf k}_{2})=
        \left[
\frac{1}{N}\sum\nolimits_{j=1}^{N}
\langle{\rm e}^{i{\bf k}_{1}\cdot{\bf R}_{j}}\rangle_{\nu}
\langle{\rm e}^{i{\bf k}_{2}\cdot{\bf R}_{j}}\rangle_{\nu}
        \right]
= \delta_{{\bf k}_1+{\bf k}_2,{\bf 0}}\;\omega({\bf k}_1).
\end{equation}
For this to be nonzero, macroscopic translational 
invariance requires ${\bf k}_1+{\bf k}_2={\bf 0}$, 
and $\omega({\bf k}_1)$ is a real-valued function. The second
moment vanishes in the isotropic fluid phase and diagnoses the
localisation of particles.  Hence the second moment (and in fact
all higher moments, as will be discussed below) serves as an order
parameter for the transition from the fluid to the amorphous solid
state.

In a different equilibrium state, denoted by $\mu$, all particles are
translated with respect to the equilibrium state $\nu$ by a
displacement ${\bf a}^{\mu}$, so that 
\begin{equation}
\langle{\rm e}^{i{\bf k}\cdot{\bf R}_{j}}\rangle_{\mu}=
{\rm e}^{i{\bf k}\cdot{\bf a}^{\mu}}
\langle{\rm e}^{i{\bf k}\cdot{\bf R}_{j}}\rangle_{\nu}\,.
\end{equation}
Due to the translational invariance of the Hamiltonian there is in
fact a manifold of symmetry-related states, generated by spatially
uniform translations. How do we have to generalise the representation
of the order parameter (\ref{naive_orderpar}), in order to take into
account the underlying translational symmetry of the Hamiltonian? 
For two equilibrium states, $\mu$ and $\rho$, the second moment 
is given by
\begin{equation}
\Omega_2^{\mu,\rho}({\bf k}_1,{\bf k}_2)=
        \left[
\frac{1}{N}\sum\nolimits_{j=1}^{N}
\langle{\rm e}^{i{\bf k}_1\cdot{\bf R}_{j}}\rangle_{\mu}
\langle{\rm e}^{i{\bf k}_2\cdot{\bf R}_{j}}\rangle_{\rho}
        \right]
  =  {\rm e}^{i{\bf k}_1\cdot({\bf a}^{\mu}-{\bf a}^{\rho}})\,
\delta_{{\bf k}_1+{\bf k}_2,{\bf 0}}\;\omega({\bf k}_1)\,.
\end{equation}
We introduce a longitudinal wavevector ${\bf k}_{\parallel}=({\bf
  k}_1+{\bf k}_2)/2$ and a transverse wavevector ${\bf k}_{\perp}=({\bf
  k}_1-{\bf k}_2)/2$ and rewrite the above expression
\begin{equation}
\label{generalOP}
\Omega_2^{\mu,\rho}({\bf k}_1,{\bf k}_2)=
{\rm e}^{i{\bf k}_{\perp}\cdot{\bf u}_{\perp}^{\mu,\rho}}
\delta_{{\bf k}_{\parallel},{\bf 0}}\; \omega({\bf k}_{\perp})
\end{equation}
in terms of a transverse displacement
${\bf u}_{\perp}^{\mu,\rho}=
{\bf a}^{\mu}-{\bf a}^{\rho}$.
Thereby, the second moment is represented by its 
\lq\lq magnitude\rq\rq\ 
$\omega({\bf k}_{\perp})$ and the generator of transverse translations 
${\rm e}^{i{\bf k}_{\perp}\cdot{\bf u}_{\perp}^{\mu,\rho}}$.

\begin{figure}[t]
  \begin{minipage}[t]{0.48\textwidth}
    \begin{center}
      \includegraphics[scale=0.3,angle=0]{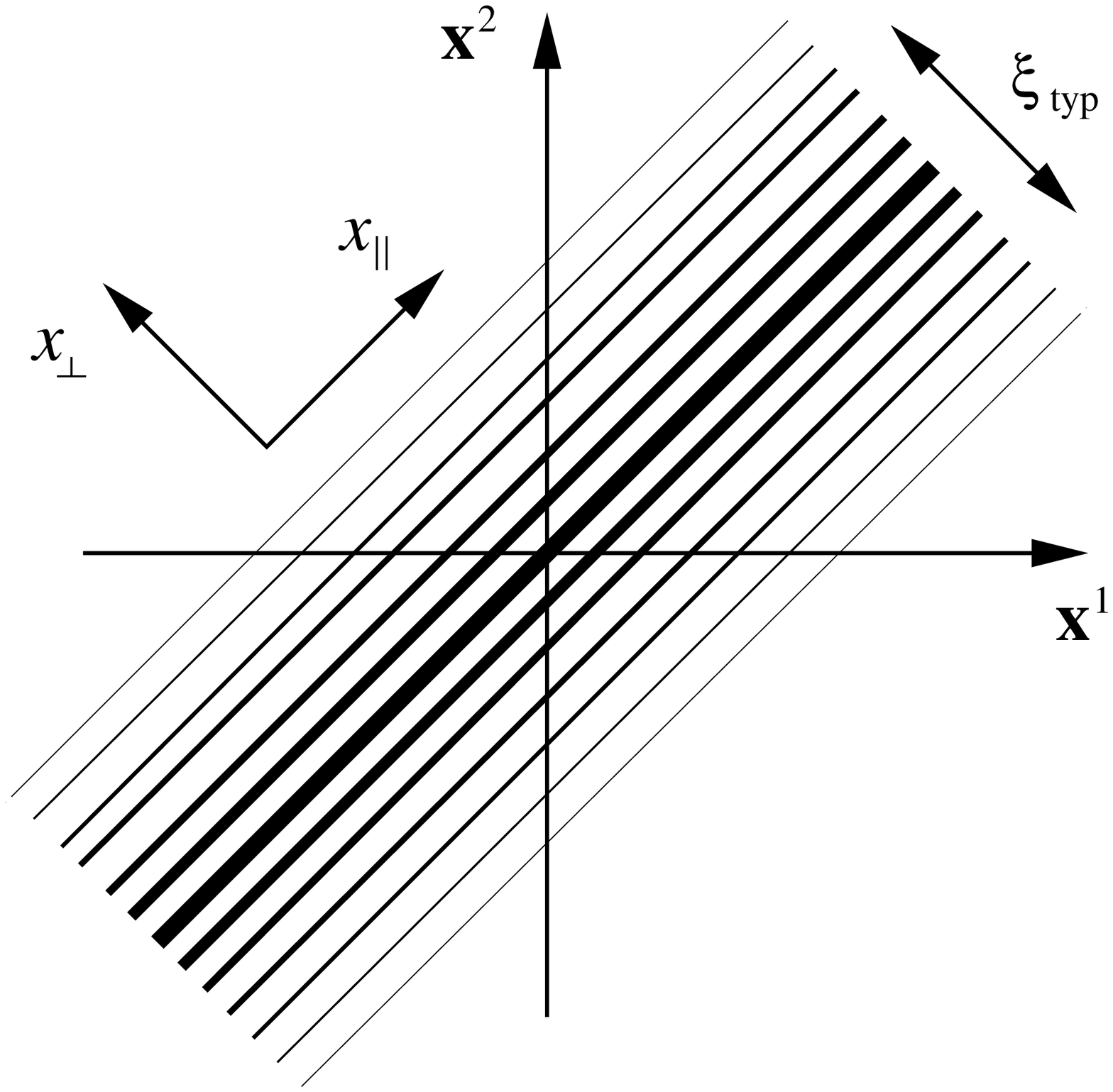}
    \end{center}
    \caption{Second moment in real space: a hill with its ridge 
    aligned in the ${\bf x}_1+{\bf x}_2$ direction.  The thickness 
    of the lines indicates the magnitude of the order parameter; 
    the position of the ridge is determined by two equilibrium states.}
      \label{fig:hill}
  \end{minipage}
  \hfill
  \begin{minipage}[t]{0.48\textwidth}
    \begin{center}
      \includegraphics[scale=0.3]{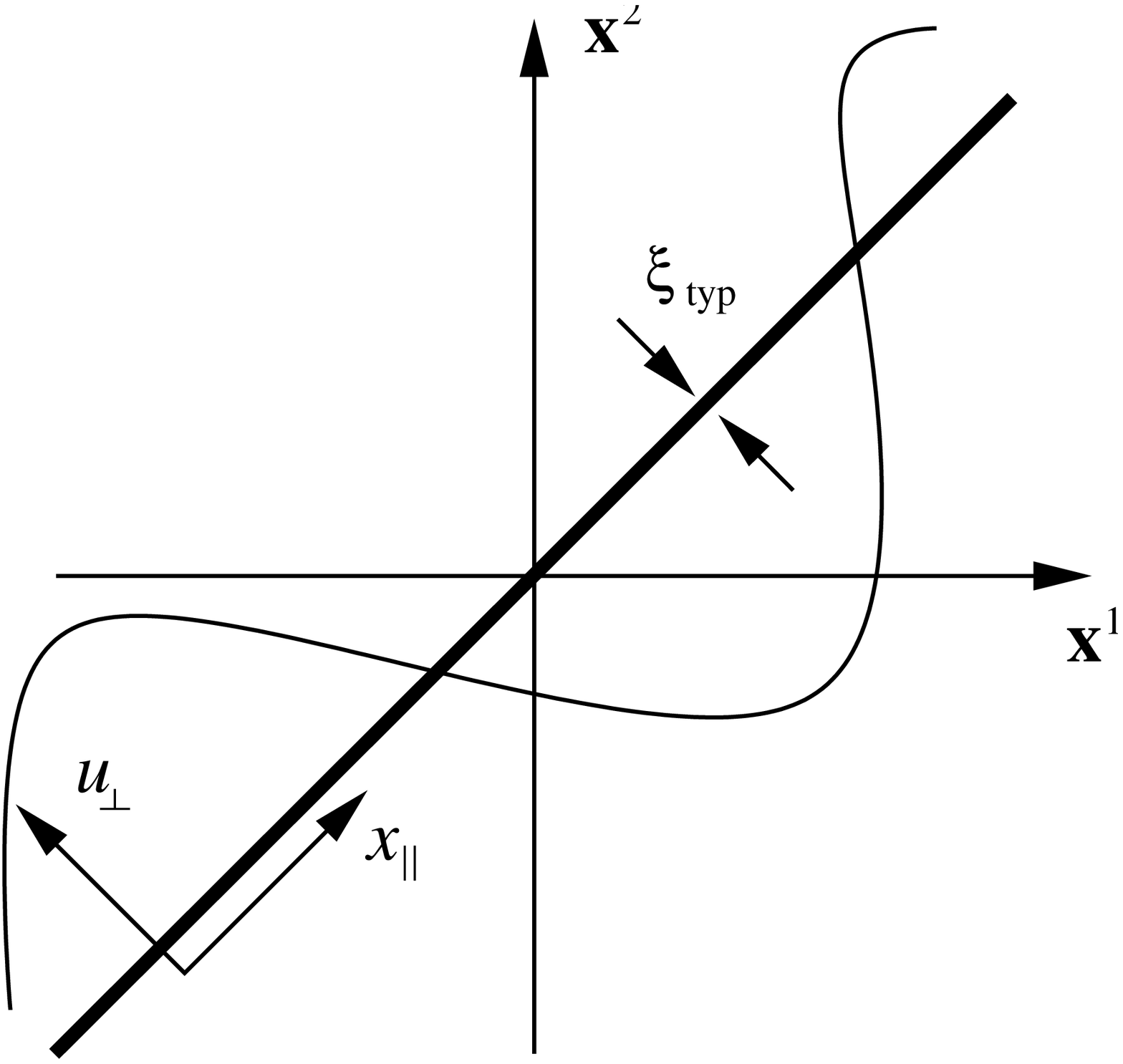}
    \end{center}
     \caption{Goldstone-distorted  state in real space:  the hill 
     is displaced perpendicularly to the ridge by an amount 
     $u_{\perp}$ that varies with position $x_{\parallel}$ along 
     the ridge.  Note that the scale of this figure is much larger 
     than that used for Fig.~\ref{fig:hill}.}
      \label{fig:ripples}
  \end{minipage}
\end{figure}

The interpretation of the order parameter is more intuitive for the
second moment as a function of real-space variables: $V {\wh
  \Omega}_2^{\mu,\rho}({\bf x}_1,{\bf x}_2)={\wh \omega}({\bf
  x}_1-{\bf x}_2-{\bf u}_{\perp}^{\mu,\rho})$.  It does not depend on
${\bf x}_{\parallel} =({\bf x}_1+{\bf x}_2)/2$, and hence can be
visualised as a rectilinear {\it hill\/}, whose position is specified
by a pair of pure states.  In a simple model
of particle localisation in $D$ dimensions, 
a fraction $Q$ of the particles are harmonically bound to random 
positions with random mean square displacements $D\,\xi_i^2$, 
giving rise to
\begin{equation}
\label{eq:simple_model_OP}
\omega({\bf k}_{\perp})= Q \;\int_0^{\infty} d\xi^2\,{\cal P}(\xi^2)
\,{\rm e}^{-\xi^2 k_{\perp}^2}.
\end{equation}
Here 
${\cal P}(\xi^{2})\equiv
\big[N_{\rm loc}^{-1}
\sum\nolimits_{j\,{\rm loc}}
\delta(\xi^{2}-\xi_{j}^{2})
\big]$ 
is the disorder-averaged distribution of squared localisation lengths 
of the $N_{\rm loc}$ localised particles.  The width of the hill is thus 
given by the typical value $\xi_{\rm typ}$ of the localisation
length. The structure in one space dimension is sketched in
Fig.~\ref{fig:hill}.

This pattern of symmetry breaking suggests that the Goldstone
excitations of an equilibrium state 
are constructed from it {\it via\/} ${\bf
  x}_{\parallel}$- dependent translations of the hill in the ${\bf
  x}_{\perp}=({\bf x}_1-{\bf x}_2)/2$ 
direction, {\it i.e.\/}, {\it ripples\/}
of the hill and its ridge (see Fig.~\ref{fig:ripples}).  
In Fourier space, the generalisation of Eq.~(\ref{generalOP}) to
inhomogeneous displacements reads
\begin{equation}
\label{eq:GDkTwo}
\Omega({\bf k}_1,{\bf k}_2)= \frac{1}{V}
\int 
d{\bf x}_{\parallel}\,
e^{i{\bf k}_{\parallel}\cdot{\bf x}_{\parallel}+
   i{\bf k}_{\perp}\cdot {\bf u}_{\perp}({\bf x}_{\parallel})}\,
\omega({\bf k}_{\perp}).
\end{equation}
The corresponding representation in real space is completely equivalent:
\begin{equation}
\label{eq:GDxTwo}
{\wh \Omega}({\bf x}_1,{\bf x}_2)= \frac{1}{V}\,
{\wh \omega}\left({\bf x}_{\perp}-{\bf u}_{\perp}({\bf x}_{\parallel})\right).
\end{equation}

\section{Goldstone fluctuations in replica space}
In the following, we apply the general concepts of Goldstone
excitations to a particular amorphous solid state---gels and
vulcanised matter. For these systems a semi-microscopic model
provides a framework for analysing the transition from the viscous
fluid to the gel, as well as the structural and thermal properties
of the gel, quantitatively. Here we use it to compute the free
energy of Goldstone fluctuations, identify it with the elastic
free energy, and thereby extract the shear modulus.

The transition from the viscous fluid to the gel is controlled by
the number of crosslinks, connecting randomly chosen pairs of
monomers. These monomers can be part of a larger prefabricated
structure, such as chains consisting of $M$ monomers or stars or
dimers {\it etc\/}. For the following, the internal structure of the
molecules is almost irrelevant; it will only determine certain
parameter values.  The average over the quenched disorder
({\it i.e.\/}~the number and location of the crosslinks) is performed with
help of the replica trick. The resulting, effectively uniform, 
theory is formulated in terms of a field ${\wh \Omega}(x)$ defined
over $(1+n)$-fold replicated space $x$, so that the argument $x$
means the collection of $(1+n)$ position $D$-vectors $\{{\bf
x}^{0},{\bf x}^{1},\ldots,{\bf x}^{n}\}$.  According to the replica
scheme of Deam and Edwards~\cite{Deam+Edwards1975}, the
disorder-averaged free energy $F$ is given by $-F/T=\lim_{n\to
0}\,\,(\rpf_{1+n}-\rpf_{1})/n\rpf_{1}$. where $T$ is the
temperature.  The replica partition function $\rpf_{1+n}$ is given
as a functional integral $\rpf_{1+n}\sim\int\fim\Omega\,e^{- {\cal
S}_{\Omega}}$ over the field $\Omega(k)$, where $k\equiv\{{\bf
k}^{0},{\bf k}^{1},\ldots,{\bf
  k}^{n}\}$ denotes wavevectors conjugate to $x$.

The expectation value of the field is given by
\begin{equation}
\Big\langle{\Omega}(k)\Big\rangle=
\left\langle
N^{-1}\sum\nolimits_{j=1}^{N}
\prod\nolimits_{\alpha=0}^{n}
{\rm e}^{i
{\bf k}^{\alpha}
\cdot
{\bf R}_{j}^{\alpha}}
\right\rangle_{{\cal S}_{\Omega}}\,.
\label{eq:PDinterpret}
\end{equation}
For this to be nonzero, 
macroscopic translational invariance requires
$\sum\nolimits_{\alpha=0}^{n}{\bf k}_{\alpha}={\bf 0}$.  The case with
only one of the $(n+1)$ wavevectors nonzero corresponds to the 
(particle number) density~(\ref{density}). 
Density fluctuations are suppressed by excluded
volume interactions, which tend to maintain homogeneity. Therefore the
theory, in its simplest version, contains only fluctuating fields for
which at least two of the $D$-component entries
in the argument $k\equiv\{{\bf k}^{0},{\bf k}^{1},\ldots,{\bf
k}^{n}\}$ are nonzero. We refer to such $D(n+1)$ dimensional
vectors as the higher-order replica sector (HRS). Fields with $l$
non-vanishing $D$-dimensional wavevectors correspond to the $l$th
moment of the local density, so in particular $l=2$ yields the
second moment discussed above. We consider a finite-dimensional 
system with a small number of nearest neighbours,
implying that fluctuations of the local density are in general
non-Gaussian and all moments need to be taken into account.

The {\it translational\/} invariance of the system is reflected in the
invariance of the effective Hamiltonian ${\cal S}_{\Omega}$ under
{\it independent translations of the replicas\/}.  
In the amorphous solid state
this symmetry is broken down to invariance under the subgroup of
common translations of all the replicas.  The manifold of 
symmetry-related classical solutions ({\it i.e.\/}~saddle
points of ${\cal S}_{\Omega}$) is given by
\begin{equation}
{\Omega}_{\rm cl}(k)=e^{i k_{\perp}\cdot u_{\perp}}\,
\delta_{{\bf k}_{\parallel},{\bf 0}}\;\,
\omega(k_{\perp}), 
\label{eq:classicalstate}
\end{equation}
in analogy to Eq.~(\ref{generalOP}).  We have decomposed 
wavevectors 
into longitudinal and transverse components with help of a complete
orthonormal basis set in replica space
$\{\ers^{\alpha}\}_{\alpha=0}^{n}$. The ``longitudinal'' direction in
$(n+1)$ dimensional replica space is the unit vector
$\ers=\sum\nolimits_{\alpha=0}^{n}\ers^{\alpha}/\sqrt{1+n}$, which is
constant in replica space.  All $D(n+1)$ dimensional vectors can be
expanded in this basis: {\it e.g.\/}, 
$k=\sum\nolimits_{\alpha=0}^{n} {\bf k}^{\alpha}\,\ers^{\alpha}$ 
and decomposed into longitudinal
($\parallel$) and transverse ($\perp$) components:
\begin{equation}
{\bf k}_{\parallel}\equiv
\sum\nolimits_{\alpha=0}^{n} {\bf k}^{\alpha}/\sqrt{1+n}
   \quad \mbox{and} \quad
k_{\perp}\equiv   k-{\bf k}_{\parallel}\,\ers\,.
\label{eq:tldecomp}
\end{equation}
Analogous expressions hold for $x$ and $u$.  Invariance under common
translations of all replicas is guaranteed in the classical solution
(\ref{eq:classicalstate}) by the Kronecker delta, which enforces the
longitudinal wavevector ${\bf k}_{\parallel}$ to be zero.  The
manifold of symmetry-related states is generated by purely transverse
displacements.

The order parameter~(\ref{eq:classicalstate}) exhibits the 
structure in replicated $x$-space discussed above for the
second moment of the local density in different equilibrium states.
The classical state $\wh{\Omega}_{\rm cl}(x)$ can be visualised as a
rectilinear hill in $x$-space with contours of constant height
oriented parallel to ${\bf x}_{\parallel}\,\ers$. Symmetry-related
classical states are obtained by translating the hill rigidly,
perpendicular to the ridge-line. Goldstone excitations from the
classical state are constructed {\it via\/} ${\bf x}_{\parallel}$-dependent
translations of the hill in the $x_{\perp}$-direction \cite{mistake},
{\it i.e.\/}, ripples of the hill and its ridge: 
\begin{subequations}
\begin{eqnarray}
\label{eq:GDkOne}
V\Omega(k)&=&
\int 
d{\bf x}_{\parallel}\,
e^{i{\bf k}_{\parallel}\cdot{\bf x}_{\parallel}+
   ik_{\perp}\cdot u_{\perp}({\bf x}_{\parallel})}\,
\omega(k_{\perp}),\\
\label{eq:GDxOne}
V\wh{\Omega}(x)&=&
{\wh\omega}\left(x_{\perp}-u_{\perp}({\bf x}_{\parallel})\right).
\end{eqnarray}
\end{subequations}

We are interested here in the low energy excitations only and, hence, 
consider Goldstone fluctuations, whose energy is expected to vanish
with a power of the inverse wavelength of the ripples. We neglect all
other fluctuations, which are expected to be massive. For consistency 
we thus should restrict ourselves to Goldstone fluctuations 
with wavelengths that are larger than a short distance cutoff $l_{<}$. 
In particular we require that the displacement 
$u_{\perp}({\bf x}_{\parallel})$ varies on length scales larger 
than the width of the interface and take $l_{<}=\xi_{\rm typ}$. 
The physical content of this restriction is that, to be meaningful,  
elastic deformations should occur on length-scales longer than the 
typical localisation length of the particles in the elastic medium. 

As discussed after Eq.~(\ref{eq:PDinterpret}), the critical fields 
of the theory reside in the HRS. Hence we want to make sure that 
the distortions of the order parameter remain in this sector.  
This is guaranteed if the distortions fulfill the incompressibility 
constraint:
$\vert\det
\left(\delta_{a,b}
+\partial_{a} u_{b}^{\alpha}
\right)\vert=1$,
which, for small amplitude distortions, reduces to
$\partial_{a} u_{a}^{\alpha}=0$.

\section{Elastic free energy and shear modulus}

For the present purposes the essential term in the
effective Hamiltonian ${\cal S}_{\Omega}$ \cite{Peng98} is the 
\lq\lq quadratic gradient\rq\rq\ term, 
\begin{equation}
\label{eq:Landaueffective}
{\cal S}_{\Omega}=
{\scriptstyle\frac{\scriptstyle 1}{\scriptstyle 2}}
Vc\sum_{k\in {\hrs}}\xi_0^{2}\,k^2\, 
\Omega(k)\,\Omega(-k)
+{\cal O}\left(k^{4}\Omega^2,\Omega^3,k^2 \Omega^3\right),
\end{equation}
where $c$ and $\xi_0$ are, respectively, the number-density and linear
size of the underlying entities being constrained and $\hrs$ indicates
that only wave vectors in the higher-order replica sector are to be
included in the summation.  Although Eq.~(\ref{eq:Landaueffective}) was
originally obtained {\it via\/} an expansion valid close to the
critical point, such a quadratic gradient term is expected to feature
in a description valid deep inside the solid state.  If one is
seeking---as we are---the dominant contribution to the elastic free
energy governing the displacement field $u_{\perp}$ ({\it i.e.\/}~the
leading-order term in $u_{\perp}$ and ${\bf k}_{\parallel}$) then it
is adequate to retain solely this quadratic gradient term.

Inserting the fluctuating field into ${\cal S}_{\Omega}$ and computing 
the {\it increase\/}, ${\cal S}_{u}$, due to the distortion 
$u_{\perp}$ ({\it i.e.\/}~the elastic free energy) gives
the following contribution, which arises solely from the quadratic
\lq\lq gradient\rq\rq\ term in ${\cal S}_{\Omega}$:
\begin{subequations}
\label{eq:elasticcouple}
\begin{eqnarray}
\label{eq:elasticenergy}
{\cal S}_{u}&=&
\frac{\stiff_{n}}{2T}\,
\int 
d{\bf x}\,
\sum_{\alpha=0}^n\;
\big(\partial_a u_b^{\alpha}({\bf x})\big)
\big(\partial_a u_b^{\alpha}({\bf x})\big),\\
\label{eq:replicastiffness}
\stiff_{n}&\equiv&
\frac{T\,c}{(1+n)^{1+\frac{D}{2}}}
\int
V^{n}\,
\dbar k_{\perp}\,
\frac{\xi_0^{2}\,k_{\perp}^{2}}{nD}\,
\omega^2(k_{\perp}).
\end{eqnarray}
\end{subequations}
Here, summation over repeated cartesian indices $a,b$ is implied and the
displacement vectors ${\bf u}^{\alpha}$ have to fulfill the constraint
$\sum\nolimits_{\alpha}{\bf u}^{\alpha}({\bf x})={\bf 0}$.

It was shown in Refs.~\cite{Castillo+Goldbart+Zippelius1994,Peng98}
that in the classical state~(\ref{eq:classicalstate})
\begin{equation}
\omega(k_{\perp})=
Q\int_{0}^{\infty}d\xi^{2}\,{\cal P}(\xi^{2})\,
e^{-\xi^{2}k_{\perp}^{2}/2},
\label{eq:mfstate}
\end{equation}
with the fraction of localised particles $Q=2a\tau/3g$
and the distribution of localisation lengths 
${\cal P}(\xi^{2})=
\big(\xi_0^{2}/a\tau\xi^{4}\big)\,
\pi\big(\xi_0^{2}/a\tau\xi^{2}\big)$,
expressed in terms of a universal classical scaling function
$\pi(\theta)$.  Here, $a\tau$ and $g$ are, respectively, 
the control parameter 
for the density of constraints and the 
nonlinear coupling constant.  
A semi-microscopic model of vulcanised
macromolecular matter yields 
$a=1/2$, 
$\tau=(\mu^{2}-\mu_{\rm c}^{2})/\mu_{\rm c}^{2}$, 
$\xi_0^{2}=M\lpc^2/2D$ and $g=1/6$, where
$\mu$ controls the mean number of constraints 
(and has critical value $\mu_{\rm c}=1$), 
$\lpc$ is the persistence length of the
macromolecules, and $M$ is the number of segments per macromolecule.
If the network is built up from single monomers, corresponding to
$M=1$, then $\lpc$ is interpreted as the length of the tether
connecting two point-like particles.

We substitute this classical value of the order parameter 
into Eq.~(\ref{eq:replicastiffness}) and take the replica 
limit $n\to 0$ to obtain the shear modulus
\begin{equation}
\stiff_{0}=
2\,T\,c\,\tau^{3}
\int d\theta\,d\theta^{\prime}\,
\frac{\pi(\theta)\,\pi(\theta^{\prime})}{\theta^{-1}+\theta^{\prime -1}}\,.
\label{eq:shearscale}
\end{equation}
The shear modulus increases continuously from zero  with the 
third power of the distance from the gel point, in agreement 
with the previous microscopic calculation 
of Ref.~\cite{Castillo+Goldbart1998-2000}.

A simple scaling argument is also consistent: The free energy density
scales with the correlation length $\zeta$ as $T \zeta^{-D}$ . This is
to be compared with the elastic free energy density of
Eq.~(\ref{eq:elasticenergy}). Since $\partial_a u_b$ is dimensionless
this implies $\stiff_{0} \sim \zeta^{-D} \sim \tau^{D\nu}$. At $D=6$ we
find agreement, which is to be expected since our calculation is based
on a Gaussian expansion around the classical solution.

\section{Destruction of localisation in 2D and quasi long range order}

We discuss the effects of Goldstone fluctuations at the level of
Gaussian fluctuations~(\ref{eq:elasticenergy}) only. The elastic
Green function is given by
\begin{equation}
{\cal G}_{ab}({\bf r})=
\int^{2\pi/l_{<}}_{2\pi/L}\dbar{{\bf k}}\,
                e^{-i{\bf k}\cdot{\bf r}}
\left(\delta_{a,b}-\hat{{\bf k}}_{a}\,\hat{{\bf k}}_{b}
\right)\frac{1}{k^2}, 
\end{equation}
where $\hat{{\bf k}}\equiv{\bf k}/\vert{\bf k}\vert$. 
Here, the integration over wave-numbers is restricted to 
$1/L<k/2\pi<1/l_{<}\,$, 
where $L$ is the linear size of the system.
The mean value of the distorted classical state, 
\begin{equation}
\label{eq:orderparameter}
\big\langle
V\Omega(k)
\big\rangle
\approx 
\int 
d{\bf x}_{\parallel}\,
e^{i{\bf k}_{\parallel}\cdot{\bf x}_{\parallel}}
\big\langle
e^{ik_{\perp}\cdot u_{\perp}({\bf x}_{\parallel})}\big\rangle_{{\cal S}_{u}}\,
\omega(k_{\perp})
=
\delta_{{\bf k}_{\parallel},{\bf 0}}\,
\exp\left(-T\,\Gamma_D\,k_{\perp}^{2}/2\stiff_{0}\right)\,
\omega(k_{\perp}), 
\end{equation}
is determined by the local Green function 
${\cal G}_{ab}({\bf 0})=\delta_{a,b}\,\Gamma_D$.  
In space dimension $D\geq 3$,
$\Gamma_D$ is finite, so that the order parameter has a finite
expectation value. The latter is reduced, due to Goldstone fluctuations, 
as one would expect: In addition to independent fluctuations of the
localised particles there are also collective fluctuations---transverse 
phonons---which effectively increase the extent of spatial
fluctuations of the localised particles. This is best seen by
recalling the classical structure for $\omega$,
Eq.~(\ref{eq:mfstate}):
\begin{equation}
\big\langle V\Omega(k) \big\rangle\approx 
\delta_{{\bf k}_{\parallel},{\bf 0}}\, 
 Q \int d\xi^2\,{\cal P}(\xi^2)\,
e^{-k_{\perp}^{2}(\xi^2+T\,\Gamma_D/\stiff_{0})/2}.
\end{equation}

Order parameter correlations are given within the Gaussian
approximation by
\begin{equation}
\big\langle
V\Omega({\bf x},k_{\perp})\,
\,V\Omega^{*}({\bf x}^{\prime},k_{\perp})
\big\rangle\approx
\omega^2(k_{\perp})\,
\exp
\left(
{-\frac{T}{\rho_{0}}
\big({\cal G}_{a b}({\bf 0})-
     {\cal G}_{a b}({\bf x}-{\bf x}^{\prime})\big)
      k_{\perp a}\cdot k_{\perp b}}
\right). 
\end{equation}
In $D=3$, the asymptotics of the elastic Green function for large
separation $r=|{\bf x}-{\bf x}^{\prime}|$ is well known, 
$ {\cal G}_{a b}({\bf r})
        \approx
(\delta_{a,b}+\hat{{\bf r}}_{a}\hat{{\bf r}}_{b}
)/8\pi r$, 
implying for the correlator of the order parameter
\begin{equation}
\big\langle
V\Omega({\bf x},k_{\perp})\,
\,V\Omega^{*}({\bf x}^{\prime},k_{\perp})
\big\rangle\approx
\omega^2(k_{\perp})\,
\exp
\left(
-\frac{T \Gamma_D k_{\perp}^2}{\rho_{0}}
+\frac{T(k_{\perp}^2+(k_{\perp}\cdot\hat{r})^2)}{8\pi\rho_0 r}
\right),
\end{equation}
where $\hat{\bf r}\equiv{\bf r}/r$. 

In two space dimensions, $\Gamma_D$ increases logarithmically with the
system size, so that the expectation value of the order 
parameter~(\ref{eq:orderparameter}) vanishes in the macroscopic limit.
Goldstone fluctuations destroy localisation in two space dimensions, in
agreement with the Hohenberg-Mermin-Wagner theorem. Nevertheless, a 
two-dimensional amorphous solid persists, which is characterised by
algebraic decay of order parameter 
correlations, {\it i.e.\/}, quasi-long-range
(random) order.  The asymptotics of the elastic Green function
for large spatial separation is given by 
${\cal G}_{a b}({\bf 0})-{\cal G}_{a b}({\bf r})\sim
{\scriptstyle\frac{\scriptstyle 1}{\scriptstyle 4\pi}}
\delta_{a,b}\,\ln(r/l_{<})$, 
implying an algebraic decay of the order-parameter correlations with a
non-universal exponent $\eta_k$: 
\begin{equation}
\big\langle
V\Omega({\bf x},k_{\perp})\,
\,V\Omega^{*}({\bf x}^{\prime},k_{\perp})
\big\rangle\sim
(r/l_{<})^{-\eta_k}\omega^2(k_{\perp})\quad \mbox{with} \quad \eta_k=
\frac{Tk_{\perp}^2}{4\pi \rho_0}\,.
\end{equation} 
This scenario is similar to the theory of two-dimensional regular 
solids, taking into account harmonic phonons only, as discussed in 
Ref.~\cite{Jancovici}.  The main difference is the absence of Bragg 
peaks at reciprocal lattice vectors. Instead, there is a divergence 
in the scattering function at ${\bf k}_{\parallel} = {\bf 0}$, as 
can be seen by integrating the correlator, above, over all 
${\bf r}={\bf x}-{\bf x}^{\prime}$\cite{Jancovici}:
$S({\bf 0},k_{\perp})=
\int d{\bf r}\, 
\langle
V\Omega({\bf r},k_{\perp})\,\,
V\Omega^{*}({\bf 0},k_{\perp})
\rangle\sim (L/\newleta)^{2-\eta_k}$.

\acknowledgments

Certain ideas concerning the nature of the vulcanization transition
in two spatial dimensions, developed in detail in the present Letter,
originated in enlightening discussions involving Horacio E.~Castillo
and Weiqun Peng;
see Refs.~\cite{GoldbartTrieste2000} and \cite{Peng+Goldbart2000}.
It is a pleasure to acknowledge these discussions, as well as
enlightening discussions with Matthew P.~A.~Fisher and Dominique Toublan.
We thank for their hospitality
the Department of Physics at the University of Colorado--Boulder (PMG)
and the Kavli Institute for Theoretical Physics at the University of
California--Santa Barbara (PMG and AZ), where some of the work reported
here was undertaken.
This work was supported in part by the National Science Foundation
under grants
NSF DMR02-05858 (SM, PMG),
NSF  PH99-07949 (PMG, AZ), 
and by the DFG through SFB~602 and Grant No.~Zi 209/6-1 (AZ).


\begin{thebibliography}{99}

\frenchspacing

\bibitem{Treloar}
  \Name{Treloar L. R. G.}
  \Book{The Physics of Rubber Elasticity}
  \Publ{Oxford University Press, Oxford}{1975}

\bibitem{Castillo+Goldbart1998-2000}
  \Name{Castillo H. E. \and Goldbart P. M.}
  \REVIEW{Phys. Rev. E}{58}{1998}{R24} and
  \REVIEW{Phys. Rev. E}{62}{2000}{8159}.

\bibitem{GoldbartAdvPhy1996}
  \Name{Goldbart P. M., Castillo H. E. \and Zippelius A.}
  \REVIEW{Adv. Phys.}{45}{1996}{393}.

\bibitem{Deam+Edwards1975}
  \Name{Deam R. T. \and Edwards S. F.}
  \REVIEW{Proc. Trans. R. Soc. Lon. Ser. A}{280}{1976}{317}.

 \bibitem{Castillo+Goldbart+Zippelius1994}
   \Name{Castillo H.E., Goldbart P.M. \and Zippelius A.}
   \REVIEW{Europhys. Lett.}{28}{1994}{519}.

\bibitem{Peng98}
  \Name{Peng W., Castillo H. E., Goldbart P.M. \and Zippelius A.}
  \REVIEW{Phys. Rev. B}{57}{1998}{893}.

\bibitem{mistake}  In our first attempts to identify Goldstone excitations
(\Name{Goldbart P. M. \and Zippelius A.}
  \REVIEW{Phys. Rev. Lett.}{71}{1993}{2256};
  \Name{Huthmann M., Rehkopf M., Zippelius A. \and Goldbart P. M.}
  \REVIEW{Phys. Rev. E}{54}{1996}{3943}) we
  allowed $u_{\perp}$ to depend not only on ${\bf x}_{\parallel}$, but
  also on $x_{\perp}$, which is incorrect in view of the above discussion.

\bibitem{Jancovici}
  \Name{Jancovici B.}
  \REVIEW{Phys. Rev. Lett.}{19}{1967}{20};
  \Name{Mikeska H. J. \and Schmidt H.}
  \REVIEW{J. Low Temp. Phys.}{2}{1970}{371}.

\bibitem{Castillo+Goldbart+Zippelius1999}
  \Name{Castillo H. E., Goldbart P. M. \and Zippelius A.}
  \REVIEW{Phys. Rev. B}{60}{1999}{14702}.

\bibitem{GoldbartTrieste2000}
See, {\it e.g.\/}, \Name{Goldbart P.M.}
\REVIEW{J. Phys. Cond. Matt.} {12}{2000}{6585}.

\bibitem{Peng+Goldbart2000}
\Name{Peng W. \and Goldbart P.M.}
\REVIEW{Phys. Rev. E}{61}{2000}{3339}.

\end{thebibliography}
\end{document}